\documentclass{amsart}
\usepackage{amssymb}
\usepackage[latin1]{inputenc}
\usepackage[swedish,english]{babel}
\usepackage{ifpdf}
\ifpdf
  \usepackage[pdftex]{graphicx}
  \usepackage{epstopdf}
\else
  \usepackage[dvips]{graphicx}
\fi
\usepackage[pdftitle={Fast sparse assembly},
  pdfauthor={Engblom/Lukarski},
  pdffitwindow=true,
  breaklinks=true,
  colorlinks=true,
  urlcolor=blue,
  linkcolor=red,
  citecolor=red,
  anchorcolor=red]{hyperref}
\usepackage[numbers,sort]{natbib}

\usepackage{caption}
\DeclareCaptionFont{white}{\color{white}}
\DeclareCaptionFormat{listing}{\colorbox{gray}{\parbox{0.8\textwidth}{#1#2#3}}}
\usepackage{listings}
\usepackage{color}

\definecolor{dkgreen}{rgb}{0,0.6,0}
\definecolor{gray}{rgb}{0.5,0.5,0.5}
\definecolor{mauve}{rgb}{0.58,0,0.82}

\newcommand{\fsparse}{\texttt{fsparse}}
\newcommand{\sparse}{\texttt{sparse}}

\newcommand{\prS}{\mbox{\texttt{prS}}}
\newcommand{\irS}{\mbox{\texttt{irS}}}
\newcommand{\jcS}{\mbox{\texttt{jcS}}}
\newcommand{\nnz}{\mbox{\texttt{nnz}}}

\newcommand{\ii}{\mbox{\texttt{ii}}}
\newcommand{\jj}{\mbox{\texttt{jj}}}
\newcommand{\jjj}{\mbox{\texttt{JJ}}}
\newcommand{\sr}{\mbox{\texttt{sr}}}
\newcommand{\len}{\mbox{\texttt{len}}}
\newcommand{\irank}{\mbox{\texttt{irank}}}
\newcommand{\irankP}{\mbox{\texttt{irankP}}}

\newcommand{\jrS}{\mbox{\texttt{jrS}}}
\newcommand{\rank}{\mbox{\texttt{rank}}}
\newcommand{\hcol}{\mbox{\texttt{hcol}}}

\numberwithin{equation}{section}
\numberwithin{table}{section}
\numberwithin{figure}{section}

\theoremstyle{plain}

\theoremstyle{definition}

\theoremstyle{remark}

\begin{document}


\title[Fast multicore sparse assembly]{Fast Matlab compatible sparse
  assembly on multicore computers}

\author[S. Engblom]{Stefan Engblom}

\author[D. Lukarski]{Dimitar Lukarski}

\address{Division of Scientific Computing \\
  Department of Information Technology \\
  Uppsala University \\
  SE-751 05 Uppsala, Sweden.}
\urladdr[(S. Engblom)]{\url{http://user.it.uu.se/~stefane}}
\email{stefane, dimitar.lukarski@it.uu.se}

\thanks{Corresponding author: S. Engblom
  (\href{mailto:stefane@it.uu.se}{stefane@it.uu.se}), telephone
  +46-18-471 27 54, fax +46-18-51 19 25.}





\date{October 23, 2015}

\begin{abstract}
  We develop and implement in this paper a fast \emph{sparse assembly}
  algorithm, the fundamental operation which creates a compressed
  matrix from raw index data. Since it is often a quite demanding and
  sometimes critical operation, it is of interest to design a highly
  efficient implementation. We show how to do this, and moreover, we
  show how our implementation can be parallelized to utilize the power
  of modern multicore computers. Our freely available code, fully
  Matlab compatible, achieves about a factor of $5\times$ in speedup
  on a typical 6-core machine and $10\times$ on a dual-socket 16-core
  machine compared to the built-in serial implementation.
\end{abstract}

\selectlanguage{english}

\maketitle


\section{Introduction}
\label{sec:intro}

The popular Matlab programming environment was originally built around
the insight that most computing applications in some way or the other
rely on storage and manipulations of one fundamental object --- the
\emph{matrix}. In the early 90s an important update was made with the
support of a \emph{sparse} storage format as presented in
\cite{matlab_sparse}. In that paper the way sparse matrices are
managed in an otherwise dense storage matrix environment is described,
including the initial creation of a sparse matrix, some basic
manipulations and operations, and fundamental matrix factorizations in
sparse format.

As a guiding principle the authors formulate the \textit{``time is
  proportional to flops''}-rule \cite[p.~334]{matlab_sparse}:
\begin{quote}
  The time required for a sparse matrix operation should be
  proportional to the number of arithmetic operations on nonzero
  quantities.
\end{quote}
The situation is somewhat different today since flops often can be
considered to be ``free'' while memory transfers are, in most cases,
the real bottlenecks of the program. With the multicore era here to
stay programs need to be threaded in order to utilize all hardware
resources efficiently. This is a non-trivial task and requires some
careful design \cite{berkeley2006}.

In this paper we consider a sole sparse operation, namely the initial
\emph{assembly} operation as implemented by the Matlab function
\sparse;
\begin{lstlisting}[language = Matlab,frame=none]
>> S = sparse(i,j,s,m,n,nzmax);
\end{lstlisting}
After the call, \texttt{S} contains a sparse representation of the
matrix defined by $S(i_{k},j_{k}) = s_{k}$ for $k$ a range of indices
pointing into the vectors $\{i,j,s\}$, \emph{and where repeated
  indices imply that the corresponding elements are to be summed
  together}. Many applications naturally lead to substantial
repetitions of indices and the implied reduction must be detected and
handled efficiently. For example, in the important case of assembly in
linear finite element methods for partial differential equations, the
resulting sparse matrix has a sparsity pattern which is identical to
that of the matrix representation of the underlying triangular or
tetrahedral mesh when viewed as a \emph{graph}. The number of
collisions during the assembly then corresponds exactly to the
connectivity of the nodes in this graph.

Since the assembly must be performed before any other matrix
operations are executed, the performance may become a bottleneck. The
typical example is for dynamic nonlinear partial differential
equations (PDEs), where re-assembly occurs many times as a numerical
time integration proceeds, including during the iterations of the
nonlinear solver. Thus, with the assembly process a quite
time-consuming operation which is repeatedly performed, it cannot
always be amortized over subsequent operations. Notably, in the truly
large case presented in \cite[\S5.1.2--5.1.3]{strong_scaling_spasm},
the performance of the sparse assembly is found to be the reason
behind the loss of strong scaling beyond a few thousands of cores.

Algorithms for sparse assembly have caught the attention also by
others. General assembly via an intermediate hashed data format is
considered in \cite{hash_sparse}, where serial performance experiment
in the PETSc library are also reported. As a follow-up on
\cite{matlab_sparse}, in \cite{matlabp_sparse} the design of sparse
matrices in Matlab*P, a kind of parallel version of Matlab, is
discussed. Unfortunately, little information about the current status
of this language is available. More recently, a ``graphBLAS''
\cite{graphBLAS} has been suggested, where one of the operations,
\texttt{BuildMatrix}, corresponds to the \sparse\ function.

As mentioned, finite element methods naturally lead to the assembly of
large sparse matrices. A stack based representation specially designed
for this application is suggested in \cite{stacked_sparse}, and is
also implemented there using a hybrid parallel programming model on a
Cray XE6. Another approach is reported in \cite{highlevel_sparse},
where the assembly of finite element sparse matrices in both Matlab
and Octave is considered using these high-level languages directly.

Using the ``time is proportional to flops''-rule as a guiding
principle - but paying close attention to memory accesses - we provide
a fast re-implementation of the \sparse\ function. The resulting
function \fsparse\ is Matlab compatible, memory efficient, and
parallelizes well on modern multicore computers. Moreover, it is well
tested and has been freely available in the public domain for quite
some time.

In \S\ref{sec:ser_fsparse} we describe in some detail the algorithm
proposed, which can be understood as an efficient index-based sorting
rule. In \S\ref{sec:par_fsparse} parallelization aspects are discussed
and performance experiments are made in \S\ref{sec:perf}, where the
memory bound character of the operation is also highlighted. In
general, with most sparse algorithms, there are not enough non-trivial
arithmetic operations to hide the format overhead and data transfer
costs \cite{sparse_challenges}. A summarizing discussion around these
issues is found in \S\ref{sec:concl}.

\subsection{Availability of software}

The code discussed in the paper is publicly available and the
performance experiments reported here can be repeated through the
Matlab-scripts we distribute. Refer to \S\ref{subsec:reproduce} for
details.

\section{A fast general algorithm for sparse assembly}
\label{sec:ser_fsparse}

In this section we lay out a fast algorithm for assembling sparse
matrices from the standard index triplet data. A description of the
problem is first offered in \S\ref{subsec:description}, where some
alternative approaches and extensions of the problem are also briefly
mentioned. The formats of input and output are detailed in
\S\ref{subsec:entry_exit} after which the algorithm is presented
stepwise in \S\ref{subsec:sparse}. A concluding complexity analysis in
\S\ref{subsec:complexity} demonstrates that the algorithm proposed has
a favorable memory access pattern without requiring large amounts of
auxiliary memory.

\subsection{Description of the problem}
\label{subsec:description}

The column compressed sparse (CCS)\footnote{Note that the abbreviation
  `CSC' is also in widespread use.} is the sparse matrix storage
format supported by Matlab but has also enjoyed a widespread use in
several other packages. Given a 4-by-4 matrix $S$ defined by
\begin{align}
  \label{eq:S}
  S &=
  \begin{pmatrix}
    10 & 0 & 0 & -2 \\
    3 & 9 & 0 & 0 \\
    0 & 7 & 8 & 7 \\
    3 & 0 & 8 & 5
  \end{pmatrix},
\end{align}
the three required CCS arrays read as follows
\begin{align*}
  \prS &= [10 \quad 3 \quad 3 \quad 9 \quad 7 \quad 8 \quad 8 \quad -2 \quad 7 \quad 5], \\
  \irS &= [0 \quad 1 \quad 3 \quad 1 \quad 2 \quad 2 \quad 3 \quad 0 \quad 2 \quad 3], \\
  \jcS &= [0 \quad 3 \quad 5 \quad 7 \quad 10],
\end{align*}
where \prS\ contains the nonzero values column-wise, \irS\ the
zero-offset row indices, and where \jcS\ points to the columns in both
\prS\ and \irS. For an $M$-by-$N$ sparse matrix we always have that
$\jcS[0] = 0$ and that $\jcS[N] = \nnz$, the total number of nonzero
elements.

In Matlab we may form a representation of $S$ by
\begin{lstlisting}[language = Matlab,frame=none]
>> s = [10 3 3 9 7 8 8 -2 7 5];
>> i = [1 2 4 2 3 3 4 1 3 4];
>> j = [1 1 1 2 2 3 3 4 4 4];
>> S = sparse(i,j,s); % size(S) = [4,4] is implicit 
\end{lstlisting}
The assembly problem, therefore, is to \emph{transform the triplet
  $(i,j,s)$ into the CCS triplet $(\irS,\jcS,\prS)$}. Generally, the
difficulties lie in that \textit{(i)} the data is unordered, and
\textit{(ii)}, the values in $s$ of equivalent pairs $(i,j)$ are to be
summed together. For example, the above matrix may also be constructed
from the triplet data
\begin{lstlisting}[language = Matlab,label =
  lst:example,caption = Sample input (running example)]
>> s = [4 4 5 7 3 5 5 4 3 4 9 7 -2];
>> i = [3 4 1 3 2 1 4 4 4 3 2 3 1];
>> j = [3 3 1 4 1 1 4 3 1 3 2 2 4];
\end{lstlisting}
Below we will use the sample input of Listing~\ref{lst:example} as a
running example to demonstrate the effects of the code snippets shown.

The complexity of the assembly operation can be bounded from above as
follows. Consider first sorting the triplet with respect to column
indices, then sorting each column with respect to rows. In a final
sweep over all columns, equivalent indices are summed together. If the
initial triplet has length $L$, then the complexity is that of sorting
and hence bounded by $L \log L$. Additionally, using an in-place
sorting algorithm, it is easy to see that the whole operation can in
fact be done in-place. In practice, sorting algorithms using auxiliary
memory are generally faster and also parallelize better.

While these are generally applicable remarks they do not take into
account the fact that the indices are bounded integers and hence can
be sorted more efficiently. In fact, the algorithm presented in
\S\ref{subsec:sparse} below can be seen as a version of the
``Distribution counting sort'' \cite[Algorithm D,
\S5.2]{TAOCP3}. Notably with this approach, speedup up to a factor of
4.8 on 8 cores was reported in \cite{parisort}.

The algorithm in \S\ref{subsec:sparse} is quite general in that it can
readily be extended to allow for various more powerful input and
output combinations. Relevant examples include the case when not all
input is available at the first call (``delayed assembly''), or when
the output must be formed in a distributed setting. Another case is
supported by the full code \fsparse\ but not detailed here; this is an
extension of the Matlab syntax which allows for row- and/or
column-indices to be counted several times as dictated by the
dimensions of the inputs.

One should keep in mind that repeated assembly can often be done
efficiently by saving various types of information between successive
calls. However, the possibility of ``quasi assembly'' is clearly very
strongly problem dependent. Thus in what follows, we remain in the
general setting.

\subsection{Format of input and output}
\label{subsec:entry_exit}

In the following we stepwise explain the algorithm by giving snippets
of actual C-code in an imagined environment which contains an
increasing number of variables. The final serial version of the code
with all pieces taken together is found in Listing~\ref{lst:sparseSeq}
in Appendix~\ref{app:ser_codes}.

In the example above the dimensions of the final matrix are implicitly
defined as the largest row- and column index. Hence as the first step,
these arrays must be parsed for the maximum values and we also
conveniently translate them into integers (recall that an array in
Matlab by default is a \texttt{double} array). Code for this
pre-processing is found in Listing~\ref{lst:getixSeq} in
Appendix~\ref{app:ser_codes} and the result is the equivalence of
Listing~\ref{lst:entry}. Note that the input index arrays remain in
unit-offset.
\begin{lstlisting}[label = lst:entry, caption = Input format]
  const double *sr;    // pointer to values
  const int *ii,*jj;   // pointers to row- and column indices (unit-offset)
  int len;             // length of arrays ii, jj, sr
  int M,N;             // sparse matrix dimensions
\end{lstlisting}

Given the input in Listing~\ref{lst:entry}, the algorithm detailed in
\S\ref{subsec:sparse} below produces an intermediate format which is
very close to the final matrix. This format is stated in
Listing~\ref{lst:exit} and contains two arrays. One of these, \jcS,
belongs to the final output and has been discussed previously. The
other one, \irank, \emph{the inverse rank}, contains information as to
how the remaining CCS-arrays \prS\ and \irS\ are to be formed from the
raw triplet data. The relation is that for $i = 0...\len-1$ (recall
that zero-offset is used here),
\begin{align}
  \label{eq:irank1}
  \irS[j] &= \ii[i]-1, \quad \mbox{where } \irank[i] = j,\\
  \label{eq:irank1b}
  \prS[j] &= \sum_{i; \; {\scriptsize \irank}[i] = j} \sr[i].
\end{align}
In plain language, $\irank[i]$ points to the final position in
$(\irS,\prS)$ for the corresponding pair $(\ii[i],\sr[i])$. Code for
finalizing the representation according to these relations is found in
Listing~\ref{lst:sparse_insertSeq} in
Appendix~\ref{app:ser_codes}. Note that as a consequence of
\eqref{eq:irank1}, $\nnz = 1+\max_{i} \irank[i]$.
\begin{lstlisting}[label = lst:exit, caption = Intermediate output format]
  int *irank;   // inverse rank array of length len
  int *jcS;     // final column pointer for sparse matrix S

  jcS = calloc(N+1,sizeof(jcS[0]));
  irank = malloc(len*sizeof(irank[0]));
\end{lstlisting}

\subsection{Index-based sparse assembly}
\label{subsec:sparse}

The task at hand is now to arrive at the intermediate output of
Listing~\ref{lst:exit}, given the parsed input of
Listing~\ref{lst:entry}. We will do this incrementally in four parts
detailed in \S\ref{subsubsec:part1}--\ref{subsubsec:part4} below. The
first part estimates the number of nonzeros per \emph{row}, the second
part constructs a rank-array which provides with the ability for a
\emph{row-wise} traversal. The most complex part of the algorithm is
the third part in which the unique row indices of each column are
found. Finally, in the fourth part the required intermediate outputs
\irank\ and \jcS\ can be determined.

\subsubsection{Part 1}
\label{subsubsec:part1}

This part builds a kind of row pointer with the same structure as
\jcS, but for rows instead of for columns. ``Kind of'' because there
is no data available to actually point into, the input still being
unordered. In Listing~\ref{lst:part1}, note also that the resulting
pointer \jrS\ ignores collisions and hence that the estimated number
of nonzeros per row is an upper bound.

\begin{lstlisting}[label = lst:part1, caption = Part 1: count rows]
  int *jrS; // accumulated row counter
  jrS = calloc(M+1,sizeof(jrS[0]));

  // count and accumulate indices to rows
  for (int i = 0; i < len; i++) jrS[ii[i]]++;
  for (int r = 2; r <= M; r++) jrS[r] += jrS[r-1];
\end{lstlisting}

\begin{proof}[Example]
  Given the arrays defined in Listing~\ref{lst:example} as inputs,
  Listing~\ref{lst:part1} produces the pointer to rows
  \begin{align*}
    \jrS &= [0 \quad 3 \quad 5 \quad 9 \quad 13].
  \end{align*}
  That collisions are ignored at this stage can be seen from the fact
  that $S$ in \eqref{eq:S} has 10 nonzero elements, whereas $\jrS$ has
  reserved space for 13 elements.
\end{proof}

\subsubsection{Part 2}
\label{subsubsec:part2}

With an upper bound on the number of nonzeros per row available it is
now straightforward to create a rank-array $\rank$ such that $i =
\rank[j]$ points to the $i$th triplet $(\ii[i],\jj[i],\sr[i])$ ordered
with respect to row indices (that is, with $\ii[i]$
non-decreasing). Hence the key feature with \rank\ is that it allows
for the data to be traversed in an ordered row-by-row fashion.

\begin{lstlisting}[label = lst:part2, caption = Part 2: build rank-array]
  int *rank; // rank-array for rows
  rank = malloc(len*sizeof(rank[0]));

  // build rank with the active use of jrS
  jrS--; /* (unit-offset in ii) */
  for (int i = 0; i < len; i++) rank[jrS[ii[i]]++] = i;
\end{lstlisting}

\begin{proof}[Example]
  Continuing with the sample input from Listing~\ref{lst:example},
  Listing~\ref{lst:part2} produces
  \begin{align*}
    \rank &= [2 \quad 5 \quad 12 \quad 4 \quad 10 \quad 0 \quad 3
            \quad 9 \quad 11 \quad 1 \quad 6 \quad 7 \quad 8], \\
    \jrS &= [* \quad 3 \quad 5 \quad 9 \quad 13 \quad 13],
  \end{align*}
  where the notation indicates that $\jrS$ is now in unit-offset. The
  defining relation is
  \begin{align*}
    \ii[\rank[\cdot]] = [1 \quad 1 \quad 1 \quad 2 \quad 2 \quad 3
    \quad 3 \quad 3 \quad 3 \quad 4 \quad 4 \quad 4 \quad 4],
  \end{align*}
  such that $\rank$ indeed provides for a row-wise traversal of the
  data.
\end{proof}

\subsubsection{Part 3}
\label{subsubsec:part3}

In this part of the algorithm, the program loops over the input and
makes each column unique with respect to row indices, building both
the index array \irank\ and the column pointer \jcS\ at the same
time. This is made feasible by the row-wise traversal of the input
data and a small cache memory for column indices.

\begin{lstlisting}[label = lst:part3, caption = Part 3: uniqueness]
  int *hcol; // cache memory for columns
  hcol = calloc(N,sizeof(hcol[0]));
  hcol--; /* (unit-offset in jj) */

  // loop over all row indices
  for (int row = 1,i = 0; row <= M; row++)
    // loop over single row
    for ( ; i < jrS[row]; i++) {
      const int ixijs = rank[i]; // index into input data triplet (ii,jj,sr)
      const int col = jj[ixijs];  // column index

      // new element?
      if (hcol[col] < row) {
	hcol[col] = row; // remembered by the row index
	jcS[col]++;        // count it
      }

      // irank keeps track of where it should go
      irank[ixijs] = jcS[col]-1;
    }

  // done: deallocate auxiliary variables
  free(++hcol);
  free(rank);
  free(++jrS);
\end{lstlisting}

\begin{proof}[Example]
  Our sample input in Listing~\ref{lst:example} yields
  \begin{align*}
    \irank &= [0 \quad 1 \quad 0 \quad 1 \quad 1 \quad 0 \quad 2 \quad
             1 \quad 2 \quad 0 \quad 0 \quad 1 \quad 0], \\
    \jcS &= [0 \quad 3 \quad 2 \quad 2 \quad 3],
  \end{align*}
  which is not very informative due to the missing final accumulation
  of indices.
\end{proof}

\subsubsection{Part 4}
\label{subsubsec:part4}

In the final part of the algorithm the column pointer \jcS\ is
finalized by an accumulating sum. Since there is a dependency between
\irank\ and \jcS, the former must be updated analogously.

\begin{lstlisting}[label = lst:part4, caption = Part 4: finalize 
    intermediate format]
  // accumulate pointer to columns
  for (int c = 2; c <= N; c++) jcS[c] += jcS[c-1];

  // irank must account for the previous accumulation
  jcS--; /* (again, unit-offset in jj) */
  for (int i = 0; i < len; i++) irank[i] += jcS[jj[i]];
  jcS++;
\end{lstlisting}

\begin{proof}[Example]
  The final part of the algorithm transforms our running example into
  \begin{align*}
    \irank &= [5 \quad 6 \quad 0 \quad 8 \quad 1 \quad 0 \quad 9 \quad
             6 \quad 2 \quad 5 \quad 3 \quad 4 \quad 7], \\
    \jcS &= [0 \quad 3 \quad 5 \quad 7 \quad 10].
  \end{align*}
  While $\rank$ is a permutation, $\irank$ is a \emph{combination} and
  has no inverse. However, if we define $\jjj$ by executing the
  assignment
  \begin{align*}
    \jjj[\irank[i]] &= \jj[i], \\
    \intertext{from $i = 0$ and upwards, then,}
    \jjj &= [1 \quad 1 \quad 1 \quad 2 \quad 2 \quad 3 \quad 3 \quad 4
           \quad 4 \quad 4].
  \end{align*}
  That is, $\irank$ has sorted the data according to columns and
  detected all collisions in the process. Note also that, as required,
  $\jjj$ is indexed by $\jcS$.
\end{proof}

\subsection{Complexity}
\label{subsec:complexity}

The assembly process clearly has a memory bound character and the
single most important complexity metric is therefore the number of
memory accesses made. Since sparse matrices are used to avoid
excessive memory use, it is also of interest to look at the amount of
working memory allocated. We estimate these both characteristics in
turn.

Thanks to the deterministic character of all loops, it is
straightforward to determine the number of memory accesses; this
amounts to little more than just counting the pointer evaluations in
Listings~\ref{lst:part1}--\ref{lst:part4}. The result is found in
Table~\ref{tab:complexity} and it shows that the number of
\emph{indirect} accesses is about $8 L$ (with $L := \len$ for
brevity), that is, the equivalence of an array of size $L$ is looped
over in random order a total of 8 times. The most common case is that
$M$ and $N$ are much smaller than $L$ such that indirect accesses to
an array of size $L$ are more expensive than to arrays of sizes $M$ or
$N$. For this situation we see that a size $L$ array is looped over
randomly a total of 3 times only.

\begin{table}
  \begin{tabular}{rrrr}
    & \#Accesses & /indirect & /size $L$\\
    \hline
    Part 1 & $2L+M$ & $L$ & 0 \\
    Part 2 & $3L$ & $2L$ & $L$ \\
    Part 3 & $5L+M$ & $4L$ & $2L$ \\
    Part 4 & $3L+N$ & $L$ & 0 \\
    \hline
    Total & $13L+2M+N$ & $8L$ & $3L$
  \end{tabular}
  \caption{Memory access complexity in
    Listings~\ref{lst:part1}--\ref{lst:part4} and in terms of $L =
    \len$. Included is the number of data accesses, the number of
    \emph{indirect} (hence possibly non-contiguous) accesses, and the
    number of indirect accesses to data arrays of size $L$ (assuming
    $L \gg M,N$).}
  \label{tab:complexity}
\end{table}

When it comes to allocated memory it is also easy to follow the
explicit allocations made by the program. The result is that the
maximal allocation will take place in one of two places. The first
candidate is in Listing~\ref{lst:part3} just after the array \hcol\
has been allocated. Here the equivalence of an integer array of size
\begin{align}
  S_{1} &=  2N+1+M+1+2L
\end{align}
has been allocated in total. A second candidate is when the final
output has been allocated. Here only the intermediate result array
\irank\ remains allocated and assuming $\mbox{\texttt{sizeof(double)}}
= 2 \times \mbox{\texttt{sizeof(int)}}$ the effective data size is
\begin{align}
  \label{eq:memory}
  S_{2} &=  N+1+3\,\nnz+L,
\end{align}
where usually $S_{2} > S_{1}$. Hence the maximal memory ever allocated
by the algorithm is generally the size of the output plus a size $L$
integer array.

As we will see in \S\ref{sec:perf} these theoretical performance
metrics do imply a respectable performance for the algorithm, with
about a factor of two times speedup compared to the built-in Matlab
version. We now proceed to parallelize the algorithm in a shared
memory environment.

\section{Parallel sparse assembly in shared memory}
\label{sec:par_fsparse}

In this section we present and analyze the threaded version of the
assembly algorithm outlined in \S\ref{sec:ser_fsparse}. The obvious
approach to parallelizing the algorithm is to evenly distribute the
input data among the cores and perform a local assembly, then finalize
the result by summing these local matrices together. We advocate
against this for two reasons: the required amount of working memory is
substantially larger with this approach and the final gather operation
is a quite complicated task in itself. Our OpenMP implementation was
developed in an incremental fashion starting from the serial
version. After several design leaps along the way the final version
achieves a competitive performance as we shall see, while still
requiring only a small amount of working memory.

\subsection{Revisited: format of input and output}

The effective input data in Listing~\ref{lst:entry} is the same in the
threaded implementation although the associated code is somewhat more
involved, see Listing~\ref{lst:getixPar} in Appendix~\ref{app:par_codes}.

The intermediate output format, however, differs considerably between
the two versions. To begin with, for the threaded version a
\emph{permuted} version \irankP\ of \irank\ is preferred. Formally,
the arrays are related through
\begin{align}
  \irankP[\rank[i]] = \irank[i],
\end{align}
for $i = 0...\len-1$. This implies that (compare
\eqref{eq:irank1}--\eqref{eq:irank1b})
\begin{align}
   \label{eq:irank2}
  \irS[j] &= \ii[k]-1, \quad \mbox{where } (\irankP[i],\rank[i]) = (j,k), \\
  \label{eq:irank2b}
  \prS[j] &= \sum_{i; \; ({\scriptsize \irankP}[i],{\scriptsize \rank}[i]) = (j,k)} \sr[k].
\end{align}

As we shall see the main benefits with this seemingly more complicated
setup is that \textit{(i)} it opens up for more parallelism in the
post-processing part, and \textit{(ii)} it simplifies the memory
access pattern in Part 3 and 4.

Code for finalizing the representation according to
\eqref{eq:irank2}--\eqref{eq:irank2b} is found in
Listing~\ref{lst:sparse_insertPar} in
Appendix~\ref{app:par_codes}. The logic for increasing the degree of
parallelism is fairly clear and builds on the fact that \rank\ allows
for a row-wise traversal; hence data can be distributed according to
row indices.

Finally, two minor details present in Listing~\ref{lst:exitP} deserve
to be mentioned. Firstly, since \rank\ is used in the post-processing
part we now need to store the row pointers \jrS\ throughout the whole
algorithm. Secondly, the algorithm is considerably streamlined when
those pointers are kept in thread-private copies. For the same reason
this design was also used for the column pointer \jcS.

\begin{lstlisting}[label = lst:exitP, caption = {Intermediate output format, parallel version}]
  // rank-array and inverse permuted rank-array
  int *rank,*irankP;
  rank = malloc(len*sizeof(rank[0]));
  irankP = malloc(len*sizeof(irankP[0]));

  const int nThreads = omp_get_max_threads();
  int **jrS; // row counters, one per thread
  jrS = malloc((nThreads+1)*sizeof(jrS[0]));
  for (int k = 0; k <= nThreads; k++) {
    jrS[k] = calloc(M+1,sizeof(jrS[k][0]));
    jrS[k]--; /* (unit-offset in ii) */
  }
  /* (final result will appear in jrS[nThreads-1]) */

  int **jcC; // column counters, one per thread
  jcS = malloc((nThreads+1)*sizeof(jcS[0]));
  for (int k = 0; k <= nThreads; k++)
    jcS[k] = calloc(N+1,sizeof(jcS[k][0]));
  /* (final result will appear in jcS[0]) */
\end{lstlisting}

\subsection{An index-based multithreaded algorithm}

The threaded version follows closely the pattern laid out in
\S\ref{sec:ser_fsparse}, the main difference being found in Parts 3
and 4 which are now merged into one parallel region.

\subsubsection{Part 1}

Counting rows in parallel is straightforward since each thread has its
own local counter. Accumulation has to be done in two steps, the last
of which is strictly serial, but which runs over a length $M$ array
only (where usually $M \ll L$). A feature with the code in
Listing~\ref{lst:part1P} is the final block in which the
thread-private pointers \jrS\ are finalized. The format used here
supports each thread to continue to process data in a fully
independent manner.

\begin{lstlisting}[label = lst:part1P, caption = {Part 1: count rows, parallel version}]
#pragma omp parallel
{
  // count local portion
  const int myId = omp_get_thread_num();
  const int istart = len*myId/nThreads;
  const int iend = len*(myId+1)/nThreads;
  for (int i = istart; i < iend; i++)
    jrS[myId+1][ii[i]]++;

#pragma omp barrier

  // accumulate jrS over the threads
#pragma omp for
  for (int r = 1; r <= M; r++)
    for (int k = 1; k < nThreads; k++)
      jrS[k+1][r] += jrS[k][r];

  // serial accumulation in jrS[0]
#pragma omp single
  for (int r = 1; r <= M; r++)
    jrS[0][r+1] += jrS[0][r]+jrS[nThreads][r];

  // determine a private jrS for each thread
#pragma omp for
  for (int r = 1; r <= M; r++)
    for (int k = 1; k < nThreads; k++)
      jrS[k][r] += jrS[0][r];
} // end parallel
\end{lstlisting}

\vspace{-1em} 

\subsubsection{Part 2}

With thread-private pointers to rows available, constructing the
rank-array is trivially parallel yet follows the logic of its serial
counterpart.

\begin{lstlisting}[label = lst:part2P, caption = {Part 2: build rank-array, parallel version}]
#pragma omp parallel
{
  // rank-array for local portion
  for (int i = istart; i < iend; i++)
    rank[jrS[myId][ii[i]]++] = i;
} // end parallel
\end{lstlisting}

\subsubsection{Part 3 and 4}

Since each thread may loop over rows independently, the double
for-loop construction in the serial code in Listing~\ref{lst:part3}
can still be used. Thanks to the modified intermediate output format,
with a permuted version \irankP\ replacing \irank, the access pattern
is actually slightly simplified. The final accumulation of \jcS\
follows closely the logic for \jrS\ in Listing~\ref{lst:part2P}.

\begin{lstlisting}[label = lst:part3_4P, caption = {Part 3+4: uniqueness and final format, parallel version}]
#pragma omp parallel
{
  int *hcol; // cache memory for columns
  hcol = calloc(N,sizeof(hcol[0]));
  hcol--; /* (unit-offset in jj) */

  const int rstart = 1+M*myId/nThreads;
  const int rend = M*(myId+1)/nThreads;
  int istart = 0;
  if (rstart > 1)
    istart = jrS[nThreads-1][rstart-1];

  // loop over segment of row indices
  for (int row = rstart,i = istart; row <= rend; row++)
    // loop over single row
    for ( ; i < jrS[nThreads-1][row]; i++) {
      const int col = jj[rank[i]]; // column index

      // new element?
      if (hcol[col] < row) {
      	hcol[col] = row;    // store row index
      	jcS[myId+1][col]++; // count it
      }

      // irankP keeps track of where it should go
      irankP[i] = jcS[myId+1][col]-1;
    }
  free(++hcol);

#pragma omp barrier

  // accumulate jcS over the threads
#pragma omp for
  for (int c = 1; c <= N; c++)
    for (int k = 1; k < nThreads; k++)
      jcS[k+1][c] += jcS[k][c];

  // serial accumulation in jcS[0]
#pragma omp single
{
  for (int c = 1; c <= N; c++)
    jcS[0][c] += jcS[0][c-1]+jcS[nThreads][c];
  jcS[0]--; /* (unit-offset in jj) */
}

  // determine a private jcS for each thread
#pragma omp for
  for (int c = 1; c <= N; c++)
    for (int k = 1; k < nThreads; k++)
      jcS[k][c] += jcS[0][c];

  // irankP must now account to these changes to jcS
  if (rend >= 1)
    for (int i = istart; i < jrS[nThreads-1][rend]; i++)
      irankP[i] += jcS[myId][jj[rank[i]]];
} // end parallel
\end{lstlisting}

\subsection{Parallel complexity}

The memory complexity of the parallel algorithm can be estimated as
before and provides us with some insight. Under the reasonable
assumption that the number of threads $p$ is small compared to the
other array sizes, for simplicity we ignore all accesses made to size
$p$ arrays. Table~\ref{tab:complexity2} lists the total number of
memory accesses performed concurrently.

Compared to Table~\ref{tab:complexity} the total number of indirect
accesses is the same in the serial and in the parallel
version. However, the number of expensive indirect accesses in size
$L$ arrays has increased from $3L$ to $4L$ which can be attributed to
the final block in Part 4 where there is now one more addressing in
the \rank-array than before. This is by design as it saves an even
more expensive final permutation of \irankP\ as well as opens up for
more parallelism in the final post-processing.

For large enough data sizes the maximal memory allocation occurs when
the final data is being allocated and is equal to the equivalent of an
integer array of size
\begin{align}
  \label{eq:memoryP}
  S_{3} &= N+1+(M+1)(p+1)+3\,\nnz+2L.
\end{align}
Compared to the serial memory footprint \eqref{eq:memory} one more
size $L$ array is required and we also see the extra allocation of
the thread-private pointer to rows \jrS.

\begin{table}
  \begin{tabular}{rrrr}
    & \#Accesses & /indirect & /size $L$\\
    \hline
    Part 1 & $2L+3Mp$ & $L$ & 0 \\
    Part 2 &  $3L$ & $2L$ & $L$ \\
    Part 3 & $5L+M$ & $3L$ & $2L$ \\
    Part 4 & $4L+3Np$ & $2L$ & $L$ \\
    \hline
    Total & $14L+3(M+N)p+M$ & $8L$ & $4L$
  \end{tabular}
  \caption{The equivalence of total number of memory accesses
    performed concurrently in $p$ threads in
    Listings~\ref{lst:part1P}--\ref{lst:part3_4P} following the
    notation in Table~\ref{tab:complexity}. By Part 4 is here meant
    the part in Listing~\ref{lst:part3_4P} which follows after the \texttt{omp
      barrier} statement.}
  \label{tab:complexity2}
\end{table}

\section{Performance Experiments}
\label{sec:perf}

In this section we present results of performance experiments of both
the proposed serial and parallel algorithms. To get some baseline
results for our index-based sorting algorithm we start by profiling
our serial version and briefly compare it with the built-in Matlab
function \sparse. The results for our threaded version are presented
in \S\ref{subsec:perf-pa}, where we look at the speedup for the
different parts of the assembly process. Although we clearly observe
the bandwidth bound character of the algorithm, the parallel
efficiency we obtain is on par with similar algorithms
\cite{parisort}. Thanks to the properties of the proposed index-based
algorithm we are able to achieve a significant speedup when compared
to the serial implementation.

\subsection{Hardware and benchmark configuration}

We performed our experiments on two different hardware platforms. The
first one is a standard workstation based on an Intel Xeon W3680 CPU
with 6 cores, 24 GB of memory and 12 MB of cache, and is denoted
`C1'. The second system is a server with a dual socket Intel Xeon
E5-2680 CPU, where each CPU has 8 cores and 20 MB of cache, 64GB of
total memory, and is denoted `C2'. On both systems we run 64 bit Linux
OS; Matlab R2014b (8.4.0.150421) on C1 and Matlab R2012b (8.0.0.783)
on C2. To avoid OS noise and caching effects, all tests were performed
40 times and the average time was determined as the arithmetic mean.

There are essentially three parameters which can vary in the input
data for the sparse assembly procedure --- the dimensions of the
output matrix, the number of nonzero elements per row of the output
matrix, and the number of collisions per row. Formally, the latter two
parameters may of course vary per row according to some empirical
distribution. For convenience we considered constant or almost
constant values only. While others have benchmarked sparse assembly
algorithms using classes of matrices ranging from highly structured to
highly unstructured cases \cite{hash_sparse}, on balance we find it
difficult to characterize this structure in a meaningful way. Some
preliminary tests performed by us using matrices from the
UF-collection \cite{UFcollection}, indicated that the actual structure
of the final matrix does not strongly influence the algorithmic
performance. We therefore ran all our tests with matrices of random
structure relying on uniform random numbers for the indices. In
Listing~\ref{DataTest} we show the generation of experimental data
according to these considerations.

\begin{lstlisting}[language = Matlab, label = DataTest, caption = Benchmark data generator]
function [ii,jj,ss,siz] = ransparse(siz,nnz_row,nrep)
% input: size, nonzeros per row, and collisions per final element
% output: row and column indices, sparse values, and size

nnz = nnz_row*siz; % number of nonzeros
ii = repmat((1:siz)',[1 nnz_row]);
jj = ceil(rand(siz,nnz_row)*siz);
ii = repmat(ii(:),[1 nrep]);
jj = repmat(jj(:),[1 nrep]); % (some jj's might be the same)

p = randperm(numel(ii));
ii = ii(p);
jj = jj(p);
ss = ones(size(ii));
\end{lstlisting}

For the experiments we focus on three different data sets. All sets
consists of 2,500,000 elements of raw input. The first data set is to
be accumulated into a sparse matrix with size 10,000 and 50 elements
per row (yielding effectively 500,000 nonzero elements in total). The
next two data sets are larger in terms of matrix size, but data set
\#2 contains less collisions and data set \#3 contains more collisions
than nonzero elements per row. The configuration is presented in
Table~\ref{tab:dataset}.

\begin{table}[h!]
  \begin{tabular}{lrrr}
    Set & matrix size & nnz & collisions \\
    \hline
    Data 1 & 10,000 & 50  & 50  \\
    Data 2 & 50,000 & 50  & 10  \\
    Data 3 & 50,000 & 10  & 50  \\
   \hline
  \end{tabular}
  \caption{Benchmark data sets. Number of nonzeros per row and number
    of collisions per nonzero element.}
  \label{tab:dataset}
\end{table}

These data could mimic different problems. For example, with finite
element methods in 3D and higher order elements, the matrices contain
a relatively large number of nonzero elements per row and the
collision pattern and resulting number of nonzero entries per row will
be fairly large (as in data set 1). The second data set could mimic
high-dimensional problems discretized with a lower order element, and
the last data set represents problems in low spatial dimension but
modeled again by higher order elements. However, the reader could map
various other scenarios from stochastic processes, data mining, and so
on to similar data sets. As a concrete example, a Laplace problem in
3D with linear Lagrange elements and discretized with tetrahedron
elements results in 12--48 collisions and about 7 nonzero elements per
row.



\subsection{Serial assembly}

\begin{figure}[!ht]
  \centering
  \includegraphics[scale=0.6]{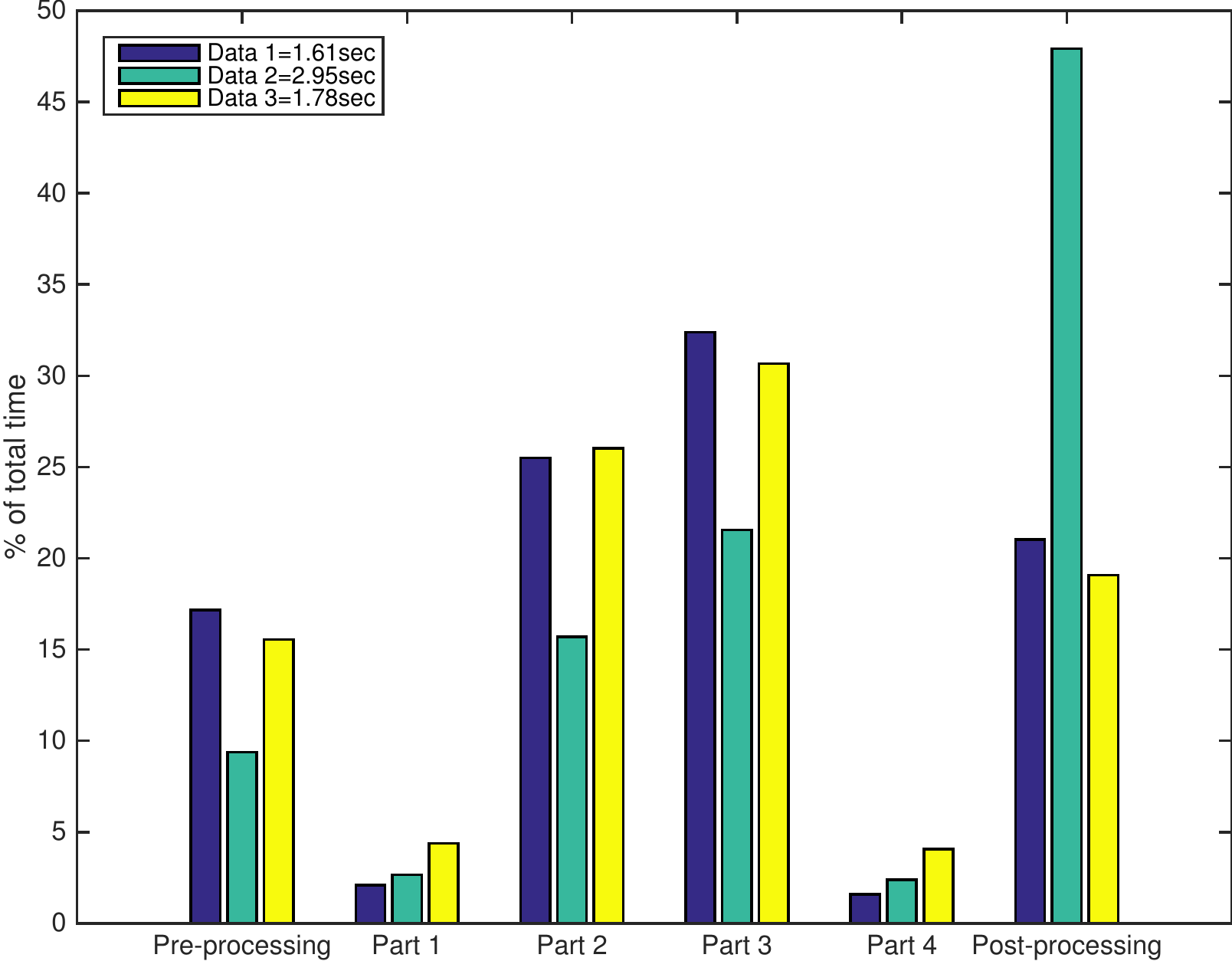}
  \caption{Total time and load distribution for pre- and
    post-processing, and for parts 1--4 (serial \fsparse, hardware
    C2).}
  \label{fig:perc-serial}
\end{figure}

In order to facilitate the understanding of the parallel behavior of
the proposed index-based sorting algorithm, we start with profiling
each part of the serial code. As expected the different data sets put
somewhat different loads at each section of the code, see
Figure~\ref{fig:perc-serial}.

%

The behavior observed in the figure can be explained fairly
intuitively as follows. If the problem contains more nonzero elements
per row (relative to the matrix size), then this puts extra pressure
on the post-processing procedure (Listing~\ref{lst:sparse_insertPar}
in Appendix~\ref{app:par_codes}) since it needs to perform more
reduction operations. As a by-product to this, naturally, the other
parts take less time in a relative sense. An important feature with
the proposed parallel algorithm is that the reduction of duplicate
elements can be done in a fully independent manner. This allows us to
perform this computation completely in parallel without any locks or
atomic operations. Counting the number of nonzero elements (Part 1)
takes under 5\% for all data sets and is proportional to the size of
the matrix and to the number of collisions per row. The operations for
achieving uniqueness (Part 2 and 3) using row-wise traversal via
\rank\ are more expensive for matrices with large number of collisions
per row. The performance of the serial \fsparse\ can of course vary
between CPUs, however, the relative time for each part is likely to
remain fairly stable with fluctuations mainly due to different cache
configurations.

Although we do not have access to Matlab's built-in \sparse\ function,
by monitor the processes we can conclude that this function is
serial. Loren Shure in a blog-post explains that \sparse\ is based on
quicksort \cite{Shure}. She remarks that quicksort has a higher
complexity compared to other algorithms such as bucket sort, but
argues that it performs fairly well in practice. For all tests
performed here, our proposed serial version outperforms Matlab's
\sparse\ convincingly, see Table~\ref{tab:matlab-serial-parallel}.


\subsection{Parallel assembly}
\label{subsec:perf-pa}

Due to the fact that essentially all operations in the sparse assembly
process are memory bounded, it is unreasonable to expect a linear
speedup even in the ideal case. The reason is that the memory bus of
the CPU can be utilized efficiently already with a single core, hence
additional memory accesses associated with an increasing number of
cores can generally not utilize the bandwidth to linear
scaling. Following the STREAM benchmark test suite \cite{McCalpin1995,
  McCalpin2007}, a very simple parallel copy function can demonstrate
this phenomenon,
\begin{lstlisting}[frame=none]
  #pragma omp parallel for
  for (int j = 0; j < N; j++) a[j] = b[j];
\end{lstlisting}
%
%
With $N = 100,000,000$ this bandwidth test shows that the OpenMP
section can speedup the copy up to $4.3\times$ on the workstation
(using 6 threads/cores) and up to $6.3\times$ (using 16 threads) on
the dual socket server. However, this is only a pure streaming test
which does not take into account the cache --- with more cores the
aggregated cache is, of course, larger.

\begin{figure}[!ht]
  \centering
  \includegraphics[scale=0.6]{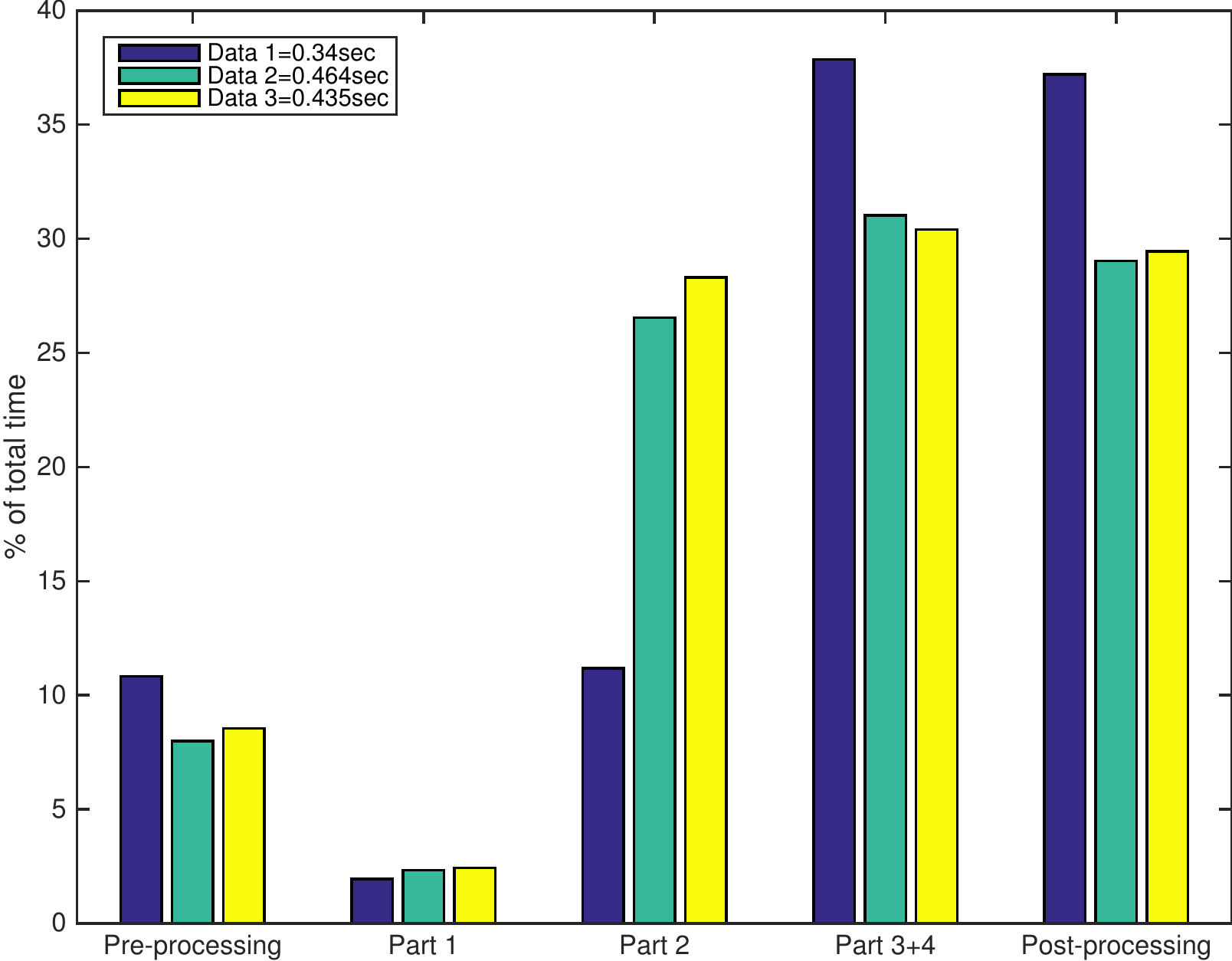}
  \caption{Total time and load distribution (parallel \fsparse, hardware
    C2).}
  \label{fig:perc-parallel}
\end{figure}

\begin{figure}[!ht]
  \centering
  \includegraphics[scale=0.6]{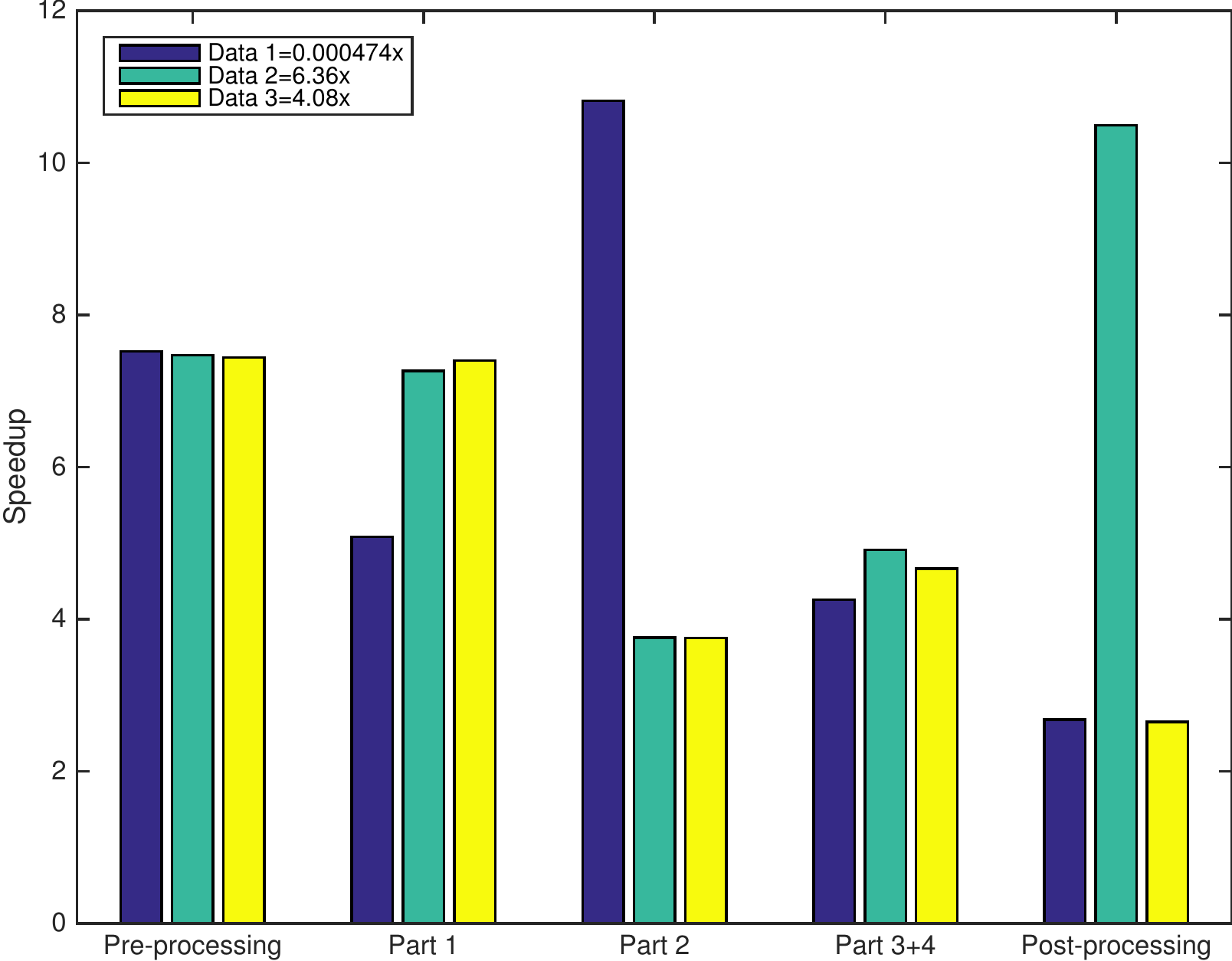}
  \caption{Parallel speedup for the different parts of \fsparse\
    (hardware C2).}
  \label{fig:parallel-perf}
\end{figure}

Before starting the actual assembly, the function determines the
maximum values of the index arrays and converts them to integers
(Listing~\ref{lst:getixPar} in Appendix~\ref{app:par_codes}). This
operation contains only contiguous memory accesses and is purely
parallel. Thus the speedup for this function does not depend on data
and is around 7$\times$ (on C2), see Figure
\ref{fig:parallel-perf}. Note that Figures~\ref{fig:perc-serial} and
\ref{fig:perc-parallel} present only the fraction of the total time,
while the actual measured speedup of course also takes into account
the real execution time.

%

For both implementations, Part 1 takes below 5\% of the total
execution time. The parallelism in this part and thus the total
speedups depend mainly on the size of the matrix, due to the fact that
all loops are performed over rows, and where the reductions are
computed element-wise per thread.

The situation is similar for the computation of the forward mapping,
the \rank-array (Part 2). The whole computation is basically a single
loop over all input entries and since the indirect mapping depends on
the number of rows in the resulting matrix, the overhead of looping
over a larger matrix reduces the speedup due to additional memory
accesses.

In the parallel version we combine Part 3 and 4 into one. Although the
memory access pattern is complex, the number of contiguous and
indirect memory accesses are proportional to the total size of the
input data arrays. Thus, the speedup factors for all test cases are in
fact similar, around 5$\times$.

All speedup factors, for all parts of the code, are presented in
Figure \ref{fig:parallel-perf}. The overall speedups comparing the
serial and parallel versions are $4.7\times$, $6.3\times$ and
$4.0\times$ on C2.

%
%

The final accumulation of the results
(Listing~\ref{lst:sparse_insertPar} in Appendix~\ref{app:par_codes})
heavily depends on the number of nonzero elements which need to be
computed. Thanks to the fact that with our approach all of the main
computations can be performed in parallel, for all data sets this
computation takes between 25--35\% of the total time. Thus, we observe
that the OpenMP implementation gives the highest speedup for the
problem with the most nonzero elements per row (normalized by the
matrix size).

Finally, we also briefly compare the full parallel \fsparse\ against
the built-in Matlab \sparse. The results are presented in
Table~\ref{tab:matlab-serial-parallel} for hardware C1 and C2. Our
implementation outperforms the Matlab version by 4--5$\times$ on C1
and 9--10$\times$ on C2.

\begin{table}[!ht]
\begin{center}
{\small
\begin{tabular}{|c|c|cc|cc|}
\hline
Data Set & MATLAB & \multicolumn{2}{c|}{Serial} & \multicolumn{2}{c|}{Parallel} \\ 
\cline{3-6}
 Hardware  & Time    & Time & Speedup   & Time & Speedup   \\
\hline
 $1$ on C1 & $3.52$ & $1.51$  & $2.33\times$ & $0.65$ & $5.39\times$ \\
 $2$ on C1 & $3.74$ & $1.87$  & $2.00\times$ & $0.83$ & $4.42\times$ \\
 $3$ on C1 & $3.49$ & $1.67$  & $2.09\times$ & $0.76$ & $4.55\times$ \\
\hline
 $1$ on C2 & $3.49$ & $1.61$  & $2.17\times$ & $0.33$ & $10.2\times$ \\
 $2$ on C2 & $4.39$ & $2.95$  & $1.49\times$ & $0.46$ & $9.71\times$ \\
 $3$ on C2 & $3.46$ & $1.78$  & $1.96\times$ & $0.43$ & $9.01\times$ \\
\hline
\end{tabular}
}
\end{center}
\caption{Overall speedup factors when comparing Matlab \sparse\ and
  \fsparse\ (serial and parallel versions).}
\label{tab:matlab-serial-parallel}
\end{table}

%

\section{Conclusions}
\label{sec:concl}

In this paper we devised an index-based implicit sorting algorithm for
the assembly of sparse matrices in CCS format given raw index-triplet
data. The algorithm was shown to be efficient in terms of memory
accesses and does not require much auxiliary memory. We also showed
how the algorithm could be modified and parallelized on multicore CPUs
with shared memory. The characteristic in terms of memory accesses for
our parallel version is remindful of the serial one and results in a
good overall performance. As shown by our experiments, compared to the
standard serial Matlab implementation, we are able to assemble a
matrix up to $10\times$ faster on a dual-socket system and about
$5\times$ faster on a 6 core system.

The approach taken in the code is a good example of how to avoid locks
by computing and storing slightly more temporary results, resulting in
a more streamlined parallel implementation and a higher efficiency.

\subsection{Reproducibility}
\label{subsec:reproduce}

Our implementation of \texttt{fsparse} as described in this paper is
available for download via the first author's
web-page\footnote{\url{http://user.it.uu.se/~stefane/freeware}}. The
code comes with a convenient Matlab mex-interface and along with the
code, automatic Matlab-scripts that repeat the numerical experiments
presented here are also distributed.

The matrix assembly functions in the
PARALUTION\footnote{\url{http://www.paralution.com}} library
(ver.0.7.0) are based on this implementation. PARALUTION is a library
for iterative sparse methods targeting multicore CPUs and
accelerators.


\section*{Acknowledgment}

The authors would like to thank master's students Aidin Dadashzadeh
and Simon Ternsjö who programmed an early version of the parallel
\fsparse\ code \cite{fsparse_report}. Their work was performed within
the project course in Computational Science at the Department of
Information Technology, Uppsala University, given by Maya Neytcheva.

The research that lead to this paper was supported by the Swedish
Research Council and carried out within the Linnaeus centre of
excellence UPMARC, Uppsala Programming for Multicore Architectures
Research Center.


\newcommand{\doi}[1]{\href{http://dx.doi.org/#1}{doi:#1}}
\newcommand{\available}[1]{Available at \url{#1}}
\newcommand{\availablet}[2]{Available at \href{#1}{#2}}

\bibliographystyle{abbrvnat}
\bibliography{el}


\appendix

\section{Additional serial codes}
\label{app:ser_codes}

The identical code producing \jj\ and $N$ has been omitted for brevity.
\begin{lstlisting}[label = lst:getixSeq, caption = Pre-processing: input of index \ii/\jj]
  const double *ival = mxGetPr(I); // Matlab input vector I
  int *ii = malloc(len*sizeof(int));
  int M = 0;
  for (int i = 0; i < len; i++) {
    // error: bad index
    if (ival[i] < 1.0 || ival[i] != ceil(ival[i])) return false;
    if ((ii[i] = ival[i]) > M) M = ival[i];
  }
\end{lstlisting}

\begin{lstlisting}[label = lst:sparse_insertSeq, caption = Post-processing: finalize CCS format]
  for (int i = 0; i < len; i++) {
    irS[irank[i]] = ii[i]-1; // switch to zero-offset
    prS[irank[i]] += sr[i];
  }
\end{lstlisting}

\begin{lstlisting}[label = lst:sparseSeq, caption = 
    \mbox{Sparse assembly, serial version}]
void sparse(const int *ii,const int *jj,const double *sr,
            int len,int M,int N)
{
  // output
  int *jcS;    // column pointer for sparse matrix S
  int *irank; // inverse rank array of length len

  int *jrS;    // accumulated "pessimistic" row counter
  int *rank;   // rank-array for rows
  int *hcol;   // cache memory for columns

  // Part 1: count and accumulate indices to rows
  jrS = calloc(M+1,sizeof(jrS[0]));
  for (int i = 0; i < len; i++) jrS[ii[i]]++;
  for (int r = 2; r <= M; r++) jrS[r] += jrS[r-1];

  // Part 2: build rank with the active use of jrS
  rank = malloc(len*sizeof(rank[0]));
  jrS--; /* (unit-offset in ii) */
  for (int i = 0; i < len; i++) rank[jrS[ii[i]]++] = i;

  /* Part 3: loop over input and make each column unique with respect
     to rowindices, building both an index vector irank and the final
     column pointer at the same time */
  jcS = calloc(N+1,sizeof(jcS[0]));
  hcol = calloc(N,sizeof(hcol[0]));
  hcol--; /* (unit-offset in jj) */
  irank = malloc(len*sizeof(irank[0]));
  for (int row = 1,i = 0; row <= M; row++)
    for ( ; i < jrS[row]; i++) {
      const int ixijs = rank[i]; // index into input data triplet (ii,jj,sr)
      const int col = jj[ixijs]; // column index

      // new element?
      if (hcol[col] < row) {
	hcol[col] = row; // remembered by the row index
	jcS[col]++;        // count it
      }

      // irank keeps track of where it should go
      irank[ixijs] = jcS[col]-1;
    }
  free(++hcol);
  free(rank);
  free(++jrS);

  // Part 4: accumulate pointer to columns
  for (int c = 2; c <= N; c++) jcS[c] += jcS[c-1];

  // irank must account for the previous accumulation
  jcS--; /* (again, unit-offset in jj) */
  for (int i = 0; i < len; i++) irank[i] += jcS[jj[i]];
  jcS++;

  /* allocate output and insert data: code not shown */

  // deallocate intermediate format
  free(irank);
}
\end{lstlisting}

\section{Additional parallel codes}
\label{app:par_codes}

\begin{lstlisting}[label = lst:getixPar, caption =  Pre-processing in parallel]
  const double *ival = mxGetPr(I); // Matlab input vector
  int *ii = malloc(len*sizeof(int));
  int M = 0;
#pragma omp parallel shared (M)
  {
    int myM = M; // local version of M
#pragma omp for
    for (int i = 0; i < len; i++) {
      if (ival[i] < 1.0 || ival[i] != ceil(ival[i]))
	ok = false; // no harm in continuing
      else if ((ii[i] = ival[i]) > myM)
	myM = ival[i];
    }

    if (M < myM)
#pragma omp critical
      // ensure nothing changed, then make the swap:
      if (M < myM) M = myM;
  } // end parallel
\end{lstlisting}

\begin{lstlisting}[label = lst:sparse_insertPar, caption = Post-processing in parallel]
#pragma omp parallel
{
  const int myId = omp_get_thread_num();
  const int rstart = 1+M*myId/nThreads;
  const int rend = M*(myId+1)/nThreads;
  int istart;
  if (rstart == 1)
    istart = 0;
  else
    istart = jrS[nThreads-1][rstart-1];

  if (rend >= 1) {
    for (int i = istart; i < jrS[nThreads-1][rend]; i++)
      irS[irankP[i]] = ii[rank[i]]-1;
    for (int i = istart; i < jrS[nThreads-1][rend]; i++)
      prS[irankP[i]] += sr[rank[i]];
  }
} // end parallel
\end{lstlisting}

\end{document}